# Results of classical and nonclassical theories of nucleation for ordering transition in $BaTiO_3$

V. Sobolev and T. Liebsch
South Dakota School of Mines and Technology, Rapid City, SD 57701

**Abstract**
The article presents a comparative analysis of the results obtained using classical and nonclassical approaches to nucleation during the ordering phase transition in barium titanate. A nonclassical nucleation model is applied to the first-order phase transition from the ferroelectric tetragonal phase to the cubic paraelectric phase in the perovskite ferroelectric barium titanate.

## I. Introduction

Mechanisms of ferroelectric (FE) phase transitions remain an important subject of investigation because they play the fundamental role in determining the formation and evolution of dipole-ordered phases in FE materials. In particular, the kinetics of a first-order FE phase transition is governed by nucleation and growth processes, through which polarized domains emerge from the parent paraelectric phase. Nucleation involves the formation of localized regions of the stable FE phase within the metastable parent phase. The formation of such nuclei requires overcoming an activation barrier associated with both the bulk free-energy difference between the two phases and the energy of the interphase boundary. Consequently, the size, energy, and stability of the critical nucleus are key quantities for describing FE phase-transition kinetics.

The properties of the critical nucleus determine the nucleation activation energy and, consequently, the rate of phase transformation. These quantities have been investigated using both theoretical models and experimental approaches see [1 - 6]. In the classical nucleation theory approximation, the nucleus assumed to have the same macroscopic properties as in the bulk of the new phase and the interphase-boundary energy is commonly assumed to be independent of temperature [6]. While this approximation may be adequate sufficiently far from the phase-transition temperature, its applicability becomes questionable in the vicinity of a FE phase transition. In this temperature range, the spontaneous polarization, dielectric susceptibility, and other thermodynamic parameters vary significantly with temperature, which can substantially modify the structure and energy of the interphase boundary. Several theoretical studies have consequently described the interphase boundary by a spatially distributed polarization profile and have considered its energy to be temperature dependent [4, 5 7, 8].

The Landau–Ginzburg–Devonshire phenomenological theory provides a convenient framework for describing the spatial variation of polarization and for calculating the properties of domain walls and interphase boundaries in FE materials (see, for example, [9-13]). Despite this capability, the temperature dependence of the interphase-boundary energy is often neglected in theoretical treatments of FE nucleation. Such an approximation may lead to an inaccurate description of the critical nucleus and nucleation barrier, particularly close to the phase-transition temperature, where the interphase-boundary structure and energy can change appreciably. Thus, the quantitative effect of the temperature dependence of the interphase-boundary energy on the critical nucleus size, nucleation barrier, and stability of the emerging FE phase requires further consideration. The temperature dependence of interphase-boundary properties in $BaTiO_3$ has recently been analyzed in [13].

The nonclassical approach to nucleation attempted to remove the above-mentioned limitations using several approaches. A phase field approach describes the temporal evolution of the non-conserved order parameter in the frames of the time-dependent Ginzburg-Landau equation, thus using the framework of non-classical nucleation theory described in ref. [14]. Phase field modeling has been used to describe the dynamics of domain growth within the stable regime of the ferroelectric phase [15-18]. Phase field model used to show that the paraelectric to FE phase transformation occurs through correlated nucleation, resulting in vortex structures of the polarization parameter [18] considered the nuclei having a sharp boundary for the polarization parameter, rather than the diffuse interphase boundary.

In the density functional theory originally developed as a nonclassical approach towards nucleation [19-21] the free energy depends on the density profile of the nuclei rather than on the radius of the new phase. A key component of density functional theory was that the capillary approximation was removed; therefore, the density of the new phase at the center of the nuclei needed not be the same as in the bulk. Moreover, the results of [19-21] show that when two or more order parameters are coupled together, one parameter dominates the pathway to the formation of the critical nucleus, with the other order parameter evolving later in the growth process.

The limited success of the above-mentioned theories in complete description of all aspects of the nucleation process led to the formation of the nonclassical two-step pathway in which at least two order parameters are necessary to describe the density and structure of the old and new phases [6]. This model agrees with Oswald's rule of stages in which: during the transformation of an unstable or metastable system into a stable one, the system prefers to reach intermediate stages having closest free energy to the initial state, rather than going directly to the stable configuration [6].

In this paper we considered the FE nucleus of the metastable (tetragonal) phase at a large distance from other nuclei, which allows neglecting the interaction between nuclei. The polarization distribution is described by the diffuse interphase boundary, which is considered as the planar interface when approaching the boundary. To this aim, each of the limiting caveats of the classical model are removed in our model by considering the following conditions:

a) The boundary between the nucleus and surrounding high symmetry matrix (diffuse boundary) is modeled by the polarization solutions for the interphase boundary [13].
b) The ratio of the interphase boundary width is compared to the value of the critical radius near $T_C$ to discuss the validity of the non-classical model.
c) The nucleus is considered to have an eccentricity value resulting in a slightly ellipsoidal shaped nucleus in nature.

The polarization inside the nucleus is still considered to approach the equilibrium value; however, our model includes a length scale comparison of the interphase boundary thickness to the critical radius value to discuss the validity of the non-classical model within the confines of the capillarity approximation. This comparison of the relative length scales of the nuclei also includes discussion of the importance of the nuclei's curvature.

This paper applies a non-classical nucleation model to the first-order paraelectric–ferroelectric (PE–FE) phase transition in $BaTiO_3$. Since order-parameter fluctuations strongly influence nucleation and the phase-coexistence range near the transition temperature, we examine nucleation from the cubic PE phase to the tetragonal FE phase. The approach is also applicable to other first-order phase transitions involving metastable states.

In ferroelectrics, electrostatic energy creates a barrier to FE nucleation. We therefore consider a diffuse interphase boundary, where polarization changes gradually between the nucleus and the surrounding high-temperature phase. This non-classical treatment lowers the energy barrier and allows isolated metastable nuclei to be modeled. The temperature dependence of the surface tension and critical nucleus radius near the Curie temperature is also evaluated, with numerical results presented for $BaTiO_3$.

The paper is organized as follows: Section II introduces our model and some results of classical nucleation theory for FE phase transitions. Section III describes the non-classical nucleation regime and the nuclei with diffuse interphase boundary. Section IV examines nuclei divided into two domains by a domain wall and contains some concluding remarks.

## II. The model and some results of classical theory of nucleation for the PE-FE phase transition.

A FE tetragonal inclusion having initially a spherical symmetry and a homogeneous polarization throughout the fiducial interior with a thin diffuse interphase boundary separating the nuclei from the surrounding matrix is considered within a paraelectric, cubic bulk crystal. The volume of the nucleus is considered large compared to the interphase boundary region allowing for a classical description of the nucleation process. The boundary between the nucleus and host matrix in our model is a diffuse interface within which the order parameter changes continuously while approaching the domain wall.

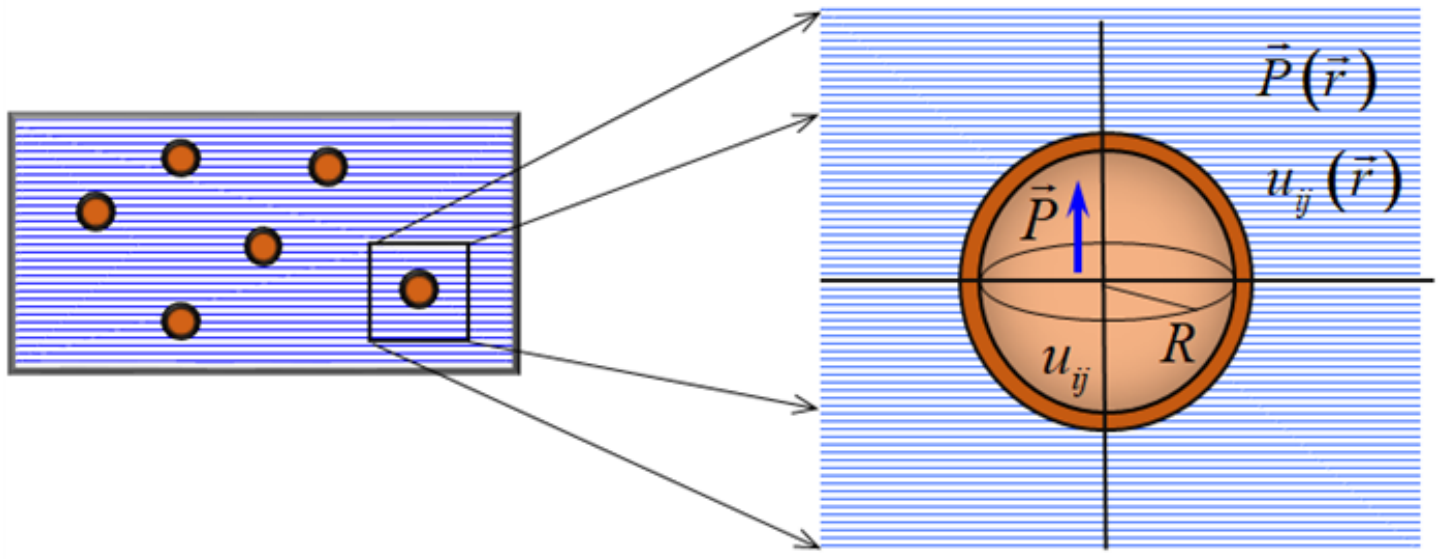

Fig. 1: Nuclei of the metastable phase appearing due to thermal fluctuations with the diffuse interphase boundary (outer ring) is shown in the insert. Polarization and strain are uniform inside the nucleus and vary outside the nucleus.

The thermodynamic potential of FE used in our studies is given by expressions [22, 23]

$$\Phi\left(P_i, P_{i,j}, u_k\right) = \Phi_P\left(P_i\right) + \Phi_{el}\left(u_k\right) + \Phi_q\left(P_i, u_k\right) + \Phi_G\left(P_{i,j}\right) \quad (1)$$

where $\Phi_P$ includes terms up to six order of polarization vector components, $\Phi_{el}$ is a potential for the elastic subsystem, $\Phi_q$ is the part describing the interaction between the polar and elastic subsystems caused by electrostriction, and the piezoelectric effect in the states without the center of symmetry, a gradient term $\Phi_G$ is needed for consideration of the inhomogeneous distributions of polarization [10, 11]. The values of coefficients in expressions (1) used in calculations were taken from ref. [22] (see also appendix A in ref. [24]).

The considered boundary is the uncharged Ising-type domain wall, forming the radial surface of the nuclei. Internal structure of the boundary between the tetragonal and cubic phases the *4mm*/*m3m* interphase boundary [13] used to describe the tetragonal nucleus in the surrounding cubic matrix has the form

$$P_z(x)=\frac{P_0}{2}\left[\frac{\sinh(x/\xi)}{\sqrt{A+\cosh^2(x/\xi)}}+1\right],\qquad A=\frac{\alpha_{111}P_0^2}{\alpha_{11}+2\alpha_{111}P_0^2},\qquad \xi=\frac{1}{P_0}\sqrt{\frac{g_{44}}{2\left(\alpha_{11}+3\alpha_{111}P_0^2\right)}},\qquad (2)$$

where $P_0 \equiv P_s(T)$ the spontaneous polarization in the bulk of the crystal far from the domain boundary, $\alpha_{11}$, $\alpha_{111}$, , $\alpha_{112}$, and $g_{44}$ are parameters of phenomenological expression (1).

The internal structure of the PE/tetragonal FE interphase boundary is presented in figure 2

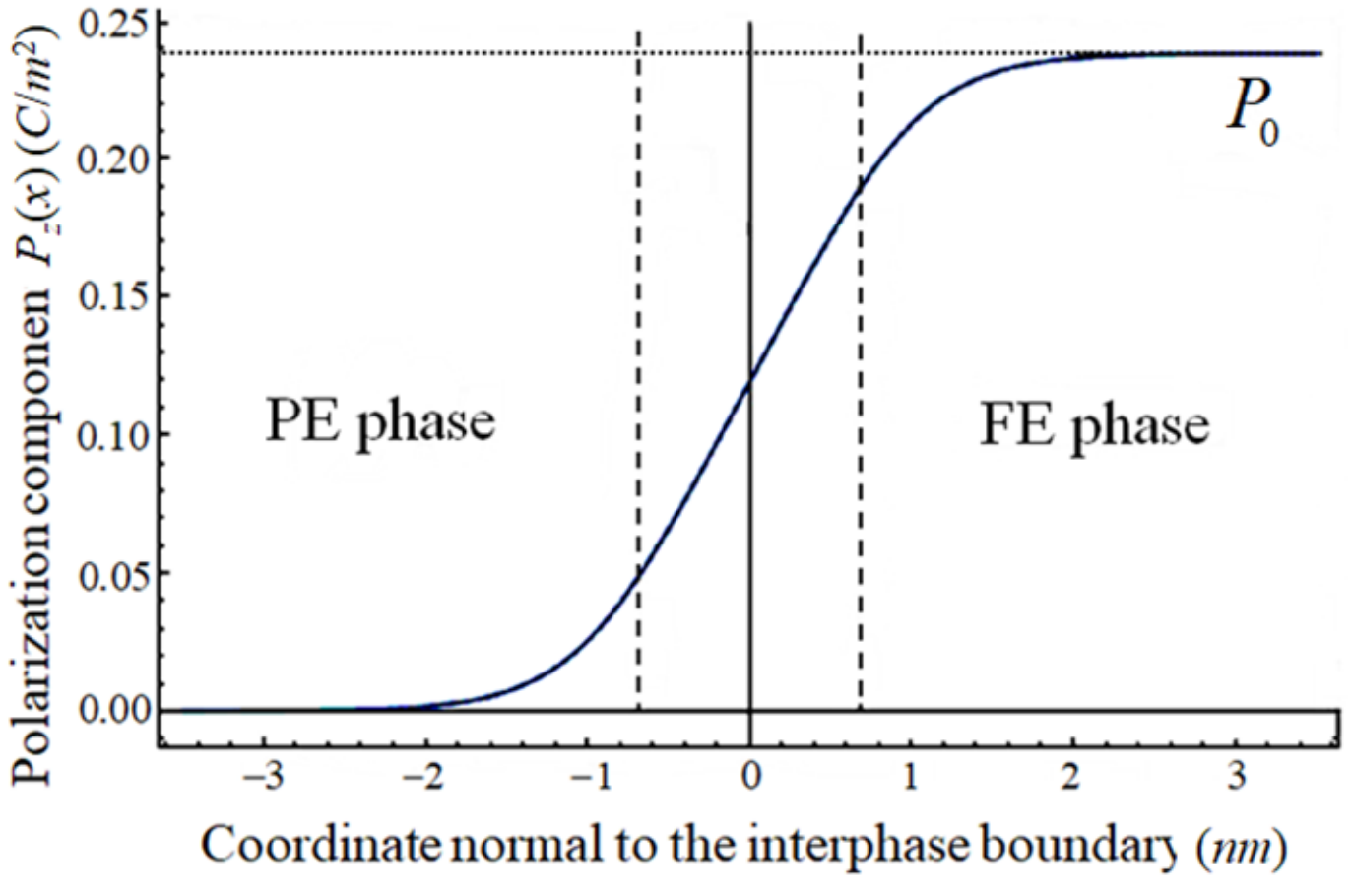

Fig. 2. Variation of polarization $P_z(x)$ inside the PE/tetragonal FE interphase boundary. The interphase boundary thickness is marked by vertical dashed lines.

The structural phase transition is accompanied by an intrinsic strain. The elastic field both inside and outside the general ellipsoidal inclusion has been considered in detail in refs. [25, 26]. Additionally, the strain solutions for a tetragonal inclusion in the host cubic matrix for the martensitic transformation was analyzed in ref. [27]. Results of refs. [25 - 27] confirm that the strain inside an ellipsoidal inclusion is uniform throughout; furthermore, an ellipsoidal inclusion can be treated as having spherical symmetry with a homogeneous strain acting as a dilation source for an isotropic medium [26]. For these reasons, this model considers a tetragonal nucleus having spherical symmetry with a uniform strain within the interior.

Here we need to comment on the statement about the insurmountable electrostatic energy barrier during the nucleation of the FE phase within the confines of classical nucleation theory (see ref. [21]). A distinction of our results is that our model considers the diffuse domain wall separating the two regions, rather than the sharp interface. The formation of nuclei increases the total energy of the material, for FE materials the electrostatic energy is the dominant contributor. The electrostatic energy of the nuclei is given by the following equation [28]

$$U_E = \frac{1}{2}\int \vec{D}\cdot\vec{E}dV = \frac{1}{2}\frac{\varepsilon_r}{\varepsilon_0 {\chi_e}^2}\int P^2 dV \ ,$$

where

$$\vec{D} = \varepsilon_0\vec{E} + \vec{P} = \varepsilon_0\varepsilon_r\vec{E} \qquad \varepsilon_r = \left(1+\chi_e\right) \qquad \vec{P} = \varepsilon_0\chi_e\vec{E} \ .$$

The dielectric constant for ferroelectric materials is much larger than unity: such that, $\varepsilon_r \gg 1 \rightarrow \varepsilon_r \approx \chi_e$. Furthermore, considering the radius of the nuclei to be large compared to the thickness of the interphase boundary, the polarization can be considered uniform throughout the volume of the nuclei. Thus,

$$U_E = \frac{1}{2}\frac{1}{\varepsilon_0\varepsilon_r}P^2\int dV = \frac{1}{2}\frac{1}{\varepsilon_0\varepsilon_r}P^2 V \qquad \rightarrow \qquad u_E = \frac{1}{2\varepsilon_0\varepsilon_r}P^2 \ . \tag{3}$$

The relative dielectric constant $\varepsilon_r$ in expression (3) is on the order of $10^3$~$10^4$ for barium titanate near the Curie temperature [29]. The elastic energy density of the nuclei (considered as elastic inclusion) derived in [25] is the same order of magnitude as the electrostatic energy density (3). Since the electrostatic and elastic energy densities are relatively the same order of magnitude, the mechanism of isolated nucleation is theoretically possible in the model considered here, and non-classical nucleation theory can be applied to the FE phase transformation.

Here we give some results of the classical nucleation theory the PE-FE PT from the cubic to tetragonal phase. The free energy density is given by

$$\Delta F = \left(\Delta\varphi_V + \Delta\varphi_S\right)\frac{4\pi}{3}r^3 + 4\pi r^2\sigma \ , \tag{4}$$

where

$$\Delta\varphi_V = \varphi_{Cubic} - \varphi_{Tetra.} = -\left(\alpha_1 P_3^2 + \tilde{\alpha}_{11} P_3^4 + \alpha_{111} P_3^6\right)$$

represents the change in the free energy density with respect to the metastable phase, $r$ is the radius of the nucleus,

$$\tilde{\alpha}_{11} = \alpha_{11} - \frac{1}{2} q_{11} Q_{11} - q_{12} Q_{12}, \quad Q_{11} = \frac{q_{11}(c_{11} + c_{12}) - 2q_{12}c_{12}}{c_{11}^2 + c_{11}c_{12} - 2c_{12}^2}, \qquad Q_{12} = \frac{q_{12}c_{11} - q_{11}c_{12}}{c_{11}^2 + c_{11}c_{12} - 2c_{12}^2},$$

where $q_{ik}$ are coefficients of the electrostriction tensor and $c_{ik}$ are elastic moduli (Voight notations are used.). The misfit strains energy density for a nucleus (as elastic inclusion) was obtained in refs. [25, 30]

$$\Delta\varphi_S = \frac{E}{15\left(1-\nu^2\right)}\left\{4u_1^2 + \left(9+5\nu\right)u_2^2 + 2\left(1+5\nu\right)u_1 u_2\right\}. \tag{5}$$

Here $E$ is Young's modulus, ν is the Poisson ratio, $u_i$ are components of them strain tensor due to the arising polarization

$$u_1 = Q_{11}P_1^2 + Q_{12}\left(P_2^2 + P_3^2\right), \qquad u_2 = Q_{11}P_2^2 + Q_{12}\left(P_1^2 + P_3^2\right), \tag{6}$$

($P_1$ and $P_2$ are zero in nucleus of tetragonal phase with $P_3$ along the tetragonal axis) and σ is the interfacial free energy density of the boundary separating the FE nucleus from host cubic matrix. The addition of strain terms to equation (1) has been considered to fit the nuclei of the new phase into the matrix of the original phase (see refs. [27, 30 - 34]). The paraelectric phase in our model is the centrosymmetric cubic phase with no heterogeneity of the polarization or strain. The classical nucleation model assumes a uniform polarization within the nucleus [35], likewise in this model the critical radius is considered much larger than the thickness of the interphase boundary, so that the polarization can be considered constant for the interior. The minimum of (4) with respect to $r$ gives the critical radius

$$R_C = \frac{2\sigma}{-\left(\Delta\varphi_V + \Delta\varphi_S\right)}$$

and then substitution into (4) gives the free energy of the barrier

$$\Delta F_{Barrier} = \frac{16\pi}{3}\frac{\sigma^3}{\left(\Delta\varphi_V + \Delta\varphi_S\right)^2}.$$

The critical radius and free energy barrier for phase nucleation given below agree with general theoretical derivations [36, 18, 30] and the derivations of ref. [35] when considering no external field.

The normalized free energy barrier given by

$$\frac{\Delta F}{\Delta F_{Barrier}} = 3\frac{r^2}{R_C^2} - 2\frac{r^3}{R_C^3}$$

describes the critical value, for which nuclei $r > R_C$ grow to form domains of the new phase; while smaller nuclei dissipate due to thermal fluctuations. It is important to note that the nucleus is considered to have sufficient size compared to the spatial fluctuations of the polarization inside the interphase boundary, so the following limiting conditions for polarization are applied

$$P_3(\infty) = 0 \qquad P_3(-\infty) = P_0 \qquad \left.{\partial P_3}/{\partial r}\right|_{r\to\pm\infty} = 0 .$$

Furthermore, the polarization is expected to change continuously throughout the interphase boundary and remain constant far from the boundary, at the center of the nucleus. The energy of the boundary between the nucleus and the surrounding matrix is denoted as $\sigma_{Shell}$ and given by the following formula.

$$\sigma_{Shell} = \int_{-\infty}^{+\infty} \left\{ \varphi(P) - \varphi(P_0) + \frac{g_{44}}{2}\left(\frac{dP}{dr}\right)^2 \right\} dr = g_{44} \int_{-\infty}^{+\infty} \left(\frac{dP}{dr}\right)^2 dr \ .$$

To determine the energy of the interphase boundary, the polarization distribution inside the cubic-tetragonal interphase boundary (2) is used and the above expression can be transformed into

$$\sigma_{Shell} = \frac{g_{44}(A+1)^2 P_0^2}{4\xi} \int_{-\infty}^{\infty} \frac{\cosh^2(s)}{\left(A+\cosh^2(s)\right)^3} ds \ ,$$

where $s$ is the coordinate normal to the interphase boundary. In the above equation, parameter $A$ is unitless, which gives the proper dimension ($J/m^2$) for the boundary surface energy. The evaluated integrand in the above formula for $\sigma_{Shell}$ gives only a scaling multiple; therefore, the temperature dependent behavior of the energy density is due to the quadratic polarization term. Evaluation of the integral gives

$$\sigma_{Shell} = \left(\frac{g_{44} P_0^2}{32\xi}\right) \mathrm{H}(A) \tag{7}$$

$$\mathrm{H}\left(A\right)=\frac{1}{A^{3/2}\sqrt{A+1}}2\left(\sqrt{A}\sqrt{A+1}(2A-1)+(4A+1)\tanh^{-1}\left(\sqrt{\frac{A}{A+1}}\right)\right).$$

As the Curie point is approached the parameter $\xi$ diverges due to the dependence on the equilibrium polarization (see formula (2)); however, the function $\mathrm{H}\left(A\right)$ remains nearly constant reaching a minimum value of 4.5 at $T_C$.

The surface energy of the interphase boundary is considerably lower than the energy of the tetragonal 180° domain wall [9]; this can be verified by numerically evaluating equation (7) to give a surface energy density of approximately 0.6 $mJ/m^2$. The energy density of the 180° domain wall having tetragonal symmetry is typically on the order of ~ 6 $mJ/m^2$ for the single-domain $BaTiO_3$ crystal [11].

The temperature dependence of the surface energy of the nuleus is presented in figure 3. Such dependece is due to the tmperature dependence on the polarization; therefore, the surface energy tends to zero at the spinodal point (point of the loos of stability) for the ferroelectric phase; which is the expected behavior from non-classical nucleation theory.

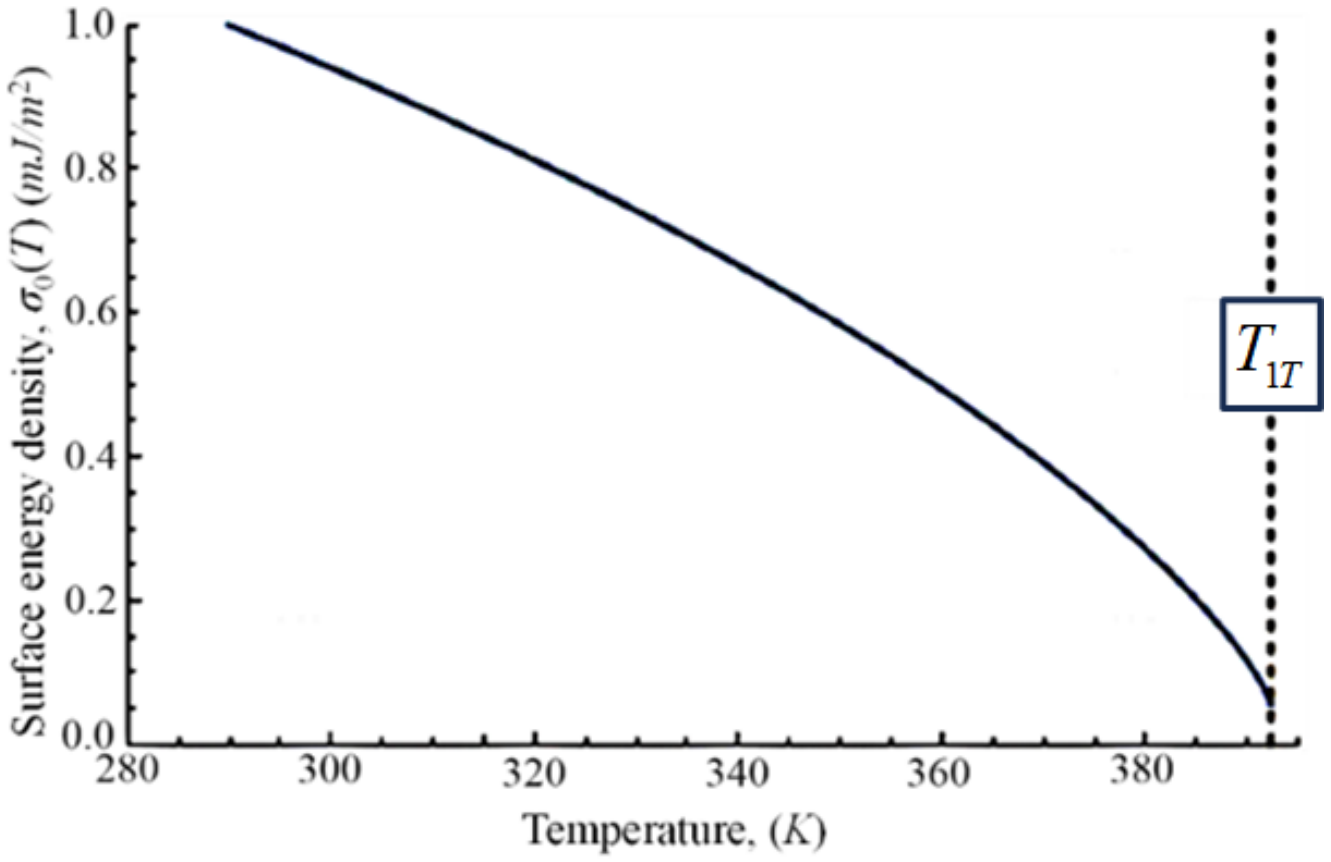

Fig.3. Temperature dependence of the surface energy density for the tetragonal nucleus. $T_{1T}$ is the temperature at which ferroelectric (tetragonal) phase loses stability at FE-PE phase transition.

Above the Curie point, the critical radius of the nucleus of the metastable ferroelectric phase diverges as the temperature is increased (as shown in figure 4). This divergence happens slightly below the point of stability loss for the tetragonal phase.

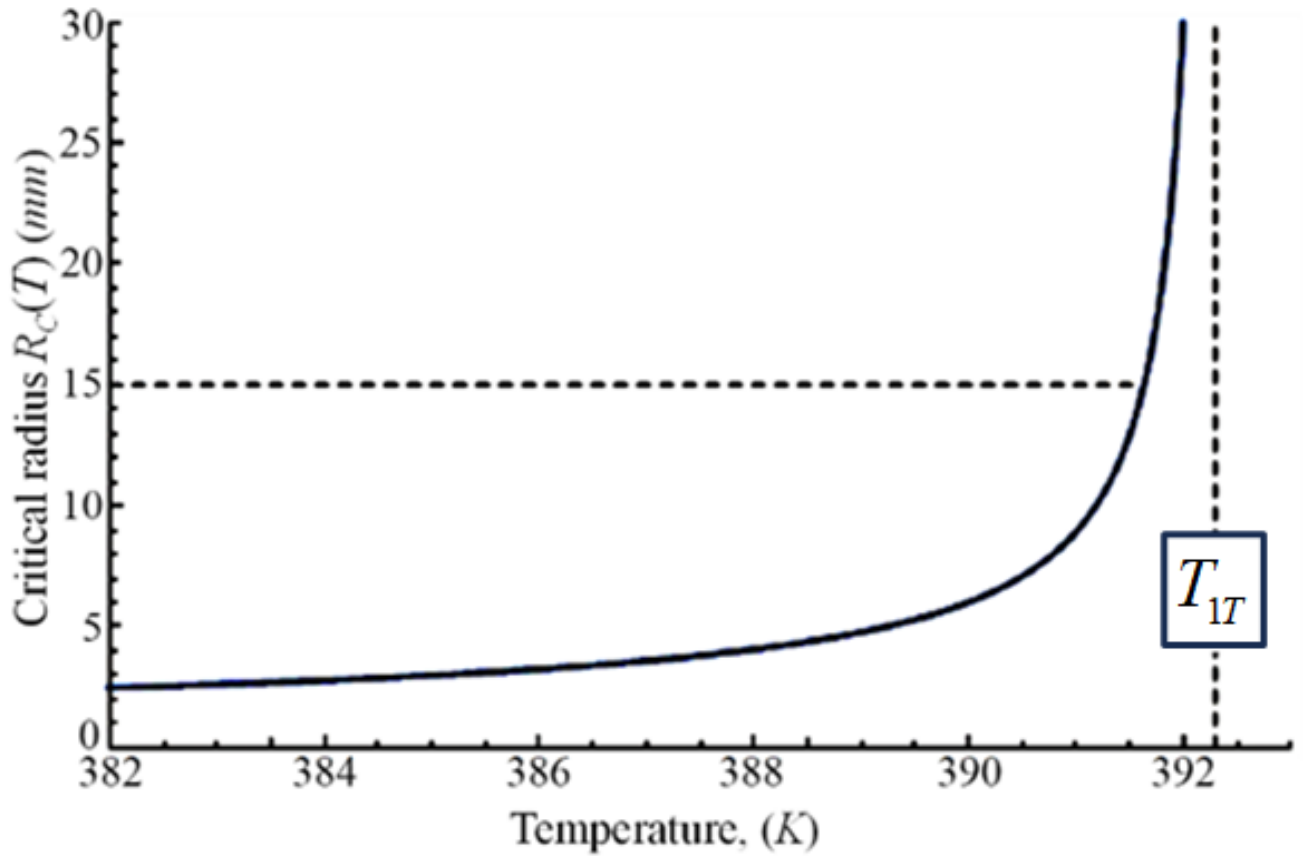

Fig. 4. Temperature dependence of the critical radius for the tetragonal nuclei inside the host cubic matrix.

By considering the additional phenomena influencing the critical radius, a more stringent temperature interval of phase stability can be defined. This smaller temperature interval more accurately matches the hysteresis behavior observed in the temperature dependence of the: dielectric constant and the lattice parameters of $BaTiO_3$. The critical radius for the formation of tetragonal nuclei is approximately 15 nm at 391 *K* (given by the horizontal dashed line in Fig. 3), which agrees with previous experimental observations [37] having a critical grain size of 30 *nm* at the Curie temperature for the cubic to tetragonal phase transformation.

If the dipole moment is aligned perpendicular to the domain wall, the total electric-dipole moment inside the nucleus can result in accumulation of charge carriers. Consideration of this specific case has been discussed in [38, 11] and results in the renormalization of the gradient coefficient $g_{44}$ proportional to a squared screening coefficient $\lambda = R_D^{-1}$ which is the reciprocal of the Debye screening radius $R_D$, with unity dielectric permeability ($\lambda^2$). Therefore, polarization solution (2) describes the component of the polarization vector in plane with the interphase boundary and the complementing solution from [9] describes the component of the polarization vector normal to the boundary; these two equations provide a complete basis for the polarization inside the nucleus. Furthermore, the polarization vector normal to the boundary creates the charged interface and the resulting surface energy density increases proportional to $g_{44} \rightarrow g_{44} + \lambda^2$.

Our results show that the critical radius for the tetragonal precipitate (Fig. 3) remains an order of magnitude larger than the width of the diffuse interphase boundary [13, 39] surrounding the nucleus for all temperatures near the first order phase transition. Furthermore, this ratio $\left(R_C / \xi\right)$ can be used to describe the lower limiting case of an initial nucleus state for which these results can be applied.

## III. Nonclassical regime of formation of the nucleus surrounded by the diffuse interphase boundary.

To consider the nucleus with the inhomogeneous polarization distribution inside the interphase boundary in the following derivations, the polarization is expressed as the field variable $P\left(\vec{r}\right)$. The free energy of the system contains the conventional Landau-Ginzburg-Devonshire terms (1); the electrostatic term contains the position-dependent polarization distribution as well as the *k*-space integral which describes static elastic energy density due to the strain distribution caused by electrostriction [40, 27, 25, 18]

$$\Delta F = \int \left\{ \Phi\left(\vec{P}\right) + g_{ijkl}\left(\frac{\partial P_i}{\partial r_j}\frac{\partial P_k}{\partial r_l}\right) - \vec{E}_{int}\cdot\vec{P} \right\} dV + \frac{1}{2}\int \frac{n_i n_j}{\varepsilon_0\varepsilon_r} P_i\left(\mathbf{r}\right) P_j\left(\mathbf{r}\right) d^3r + \frac{1}{2}\int \frac{d^3k}{\left(2\pi\right)^3}\left[K_{ijkl}\tilde{u}_{ij}\tilde{u}^*_{kl}\right] \quad (8)$$

where

$$K_{ijkl} = C_{ijkl} - n_p C_{ijpq}\Omega_{qr}C_{klrs}n_s, \quad (9)$$

$\tilde{u}^*_{ik}$ denotes complex conjugate of the Fourier transform of the strain tensor $u_{ik}$, $\Omega_{ij}\left(\vec{n}\right)$ is proportional to the Fourier transform of the Green-function of anisotropic elasticity and is defined as the inverse tensor of $\Omega^{-1}_{il}\left(\vec{n}\right) = C_{ijkl}n_j n_k$, where $C_{ijkl}$ is the elastic modulus tensor, $\vec{n} = \vec{k}/k$ [27]. It should be noted that $E_{int}$ is the internal depolarizing electric field due to the single domain inside the nucleus. The contributions of the homogeneous portion of the free energy (8) given by the first integrand to the temperature dependences of energy densities of homogeneous phases as well as energy densities of interphase boundaries has been analyzed in ref. [14, 40]; therefore, only the following two integrals need to be evaluated to describe the entire free energy of the system.

$$U_{Electro} = \frac{1}{2\varepsilon_0\varepsilon_r}\int \left|\mathbf{n}\cdot\mathbf{P}\left(\mathbf{r}\right)\right|^2 d^3r \qquad U_{Elastic} = \frac{1}{2}\int \frac{d^3k}{\left(2\pi\right)^3} K_{ijkl}\tilde{u}_{ij}\tilde{u}^*_{kl}. \quad (10)$$

An important distinction of our model is that we consider the diffuse boundary surrounding the nucleus; therefore, we do not utilize the shape function for the inhomogeneous polarization distribution (as seen in ref. [27, 18], for example) but rather use the phase field solution to describe the position-dependence of the polarization vector. Formula (2) describes the distribution of polarization inside the *4mm/m3m* interphase boundary separating the nucleus from the host matrix. When considering the finite dimensions of the metastable nuclei, the polarization is only non-zero inside the nucleus giving the following boundary conditions.

$$P(r)\big|_{r\to\infty}=0 \quad , \quad \frac{dP}{dr}\bigg|_{r\to\infty}=0 \quad , \quad \langle P(r)\rangle=0 \qquad (11)$$

Due to the presence of the diffuse boundary, the polarization within the nucleus is modeled to be dependent on the radius. Furthermore, polarization distribution must follow the behavior of solution (2) (given below for reference) in the vicinity of the interphase boundary. This distribution written in the spherical coordinate system with the origine in the nucleus center i.e. $P(x)\big|_{r=\sqrt{x^2+y^2+z^2}}\to P(r)$, acquire the form

$$P(r)=\frac{P_0}{2}\left(\frac{\sinh\left((R-r)/\xi\right)}{\sqrt{A+\cosh^2\left((R-r)/\xi\right)}}+1\right)$$

where now $R$ is the maximal radius of the nucleus is at the center of the interphase boundary. Thus, the above formula shows that the polarization within the nucleus is dependent purely on the radius and gives the polarization distribution for the tetragonal nucleus inside the cubic matrix satisfying the boundary conditions (11).

The spontaneous polarization, critical radius of the nucleus and even the trigonometric deviation parameter $A$ are all functions of temperature. A three-dimensional graphical representation of the function

$$P(r,T)=\frac{P_0(T)}{2}\left(\frac{\sinh\left[\left(R_C(T)-r\right)/\xi(T)\right]}{\sqrt{A(T)+\cosh^2\left[\left(R_C(T)-r\right)/\xi(T)\right]}}+1\right) \qquad (12)$$

is presented in figure 5.

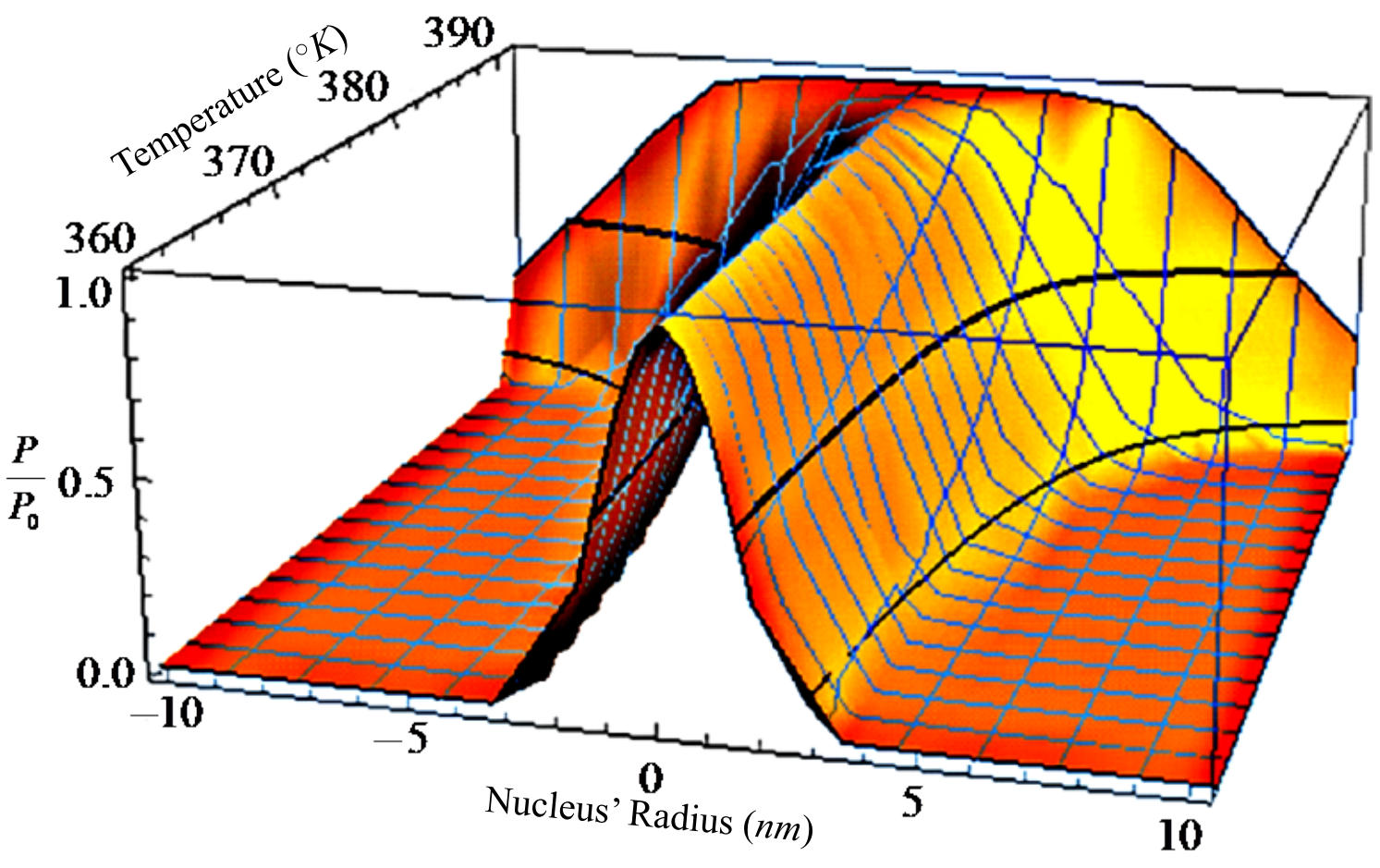

Fig. 5. Temperature and position dependence of the polarization for the nucleus having the critical radius, for a tetragonal nucleus inside the host cubic matrix.

The heavier black line shows the temperature dependence of the critical radius, evaluated at the average domain boundary width, i.e. $P/P_0 = 0.5$. The thinner black line displays $2R_C(T)$ which defines the entire region of polarization fluctuations in the vicinity of the interphase boundary. The critical radius of the nucleus represents the minimum size for the metastable phase to continue to grow. As shown in figure 4, the critical radius diverges to infinity as the Curie temperature is approached, because the metastable (polar) phase is losing stability; this results in the radius of the nuclei widening as temperature increases. In spherical coordinate system expression for $U_{Electro}$ from (10) acquires the form

$$U_{Electro} = \frac{1}{2\varepsilon_0\varepsilon_r}\int_0^{2\pi}\int_0^{\pi}\sin\theta d\phi d\theta\int_0^{R_C}\left|P(r)\right|^2 r^2 dr = \frac{\pi P_0^2}{2\varepsilon_0\varepsilon_r}\int_0^{R_C}\left|\frac{\sinh\left[(R_C-r)/\xi\right]}{\sqrt{A+\cosh^2\left[(R_C-r)/\xi\right]}}+1\right|^2 r^2 dr\,.$$

Evaluation of the above integral gives the following result

$$U_{Electro} = \frac{P_0^2}{2\varepsilon_0\varepsilon_r}\left(\frac{4\pi}{3}R_C^3\right) \quad \text{provided that} \quad \frac{R_C}{\xi} > 1 \tag{13}$$

The contribution of static elastic energy of the nucleus has been thoroughly analyzed (see ref. [25 - 37, 33, 18] for example) and is presented here for the sake of completeness. Expression from (10) is modified

$$U_{Elastic} = \frac{1}{2}\int\frac{d^3k}{(2\pi)^3}K_{ijkl}\tilde{u}_{ij}\tilde{u}_{kl}^{*}\left|\theta(\vec{k})\right|^2,$$

where the information on the shape and volume of inclusions is included in the Fourier transform of the shape function [27].

$$\left|\theta\left(\vec{k}\right)\right|^2 = \left|\int e^{-i\vec{k}\cdot\vec{r}}dV\right|^2 = V^2\left|3\frac{\sin\left(f\left(\vec{k}\right)\right)-f\left(\vec{k}\right)\cos\left(f\left(\vec{k}\right)\right)}{\left[f\left(\vec{k}\right)\right]^3}\right|^2$$

In the above equation the function $f\left(\vec{k}\right)=\sqrt{L_{ij}k_ik_j}$, where $L_{ij}$ is the tensor inverse to $\left(\hat{\mathbf{L}}\right)_{ij}$ that determines the standard form of the equation for the ellipsoid surface by $\left(L^{-1}\right)_{ij}r_ir_j=1$. The eigenvalues of this tensor are the squared values of the ellipsoid axis. As it was shown in ref. [27] due to the identity

$$V \equiv \int\frac{d^3k}{\left(2\pi\right)^3}\left|\theta\left(\mathbf{k}\right)\right|^2 = \int_0^{2\pi}\int_0^{\pi}\int_0^{\infty}\frac{k^2\sin\theta}{\left(2\pi\right)^3}d\phi d\theta dk\left|3V\frac{\sin\left(f\left(\vec{k}\right)\right)-f\left(\vec{k}\right)\cos\left(f\left(\vec{k}\right)\right)}{\left[f\left(\vec{k}\right)\right]^3}\right|^2. \quad (14)$$

One can estimate the lower limit for the elastic energy as

$$U_{Elastic}=\frac{1}{2}\int\frac{d^3k}{\left(2\pi\right)^3}K_{ijkl}\tilde{u}_{ij}\tilde{u}_{kl}^*\left|\theta\left(\vec{k}\right)\right|^2 \geq \frac{V}{2}\left[\min\left(K_{ijlk}u_{ij}^0u_{kl}^0\right)\right].$$

As was demonstrated in refs. [25, 27], ellipsoidal inclusions are unique in that they can be reduced to a sphere having the same volume by homogeneous space deformation. Thus, for the case of spherical symmetry having radius *R*, the function $f\left(\vec{k}\right)=kR$, which simplifies (14)

$$\frac{1}{4\pi V}=\int_0^{\infty}\frac{1}{\left(2\pi\right)^3}k^2dk\left|3\frac{\sin\left(kR\right)-kR\cos\left(kR\right)}{\left(kR\right)^3}\right|^2.$$

As shown in refs. [25, 27] the strain inside an ellipsoidal inclusion is uniform within the nucleus and only depends on the shape. Therefore, the elastic energy of the nucleus with the tetragonal distortions due to electro striction effects is given by (5) and after substitutions of $\varepsilon_1=Q_{12}P^2$, $\varepsilon_3=Q_{11}P^2$ has the form [18]

$$U_{Elastic}=\frac{EP^4}{15\left(1-\nu^2\right)}\left[\left(9+5\nu\right)Q_{12}^2+4Q_{11}^2+2\left(1+5\nu\right)Q_{11}Q_{12}\right]V$$

corresponding to the part of expression for the free energy given by last integral in (8).

## IV. Analysis of critical radius for nucleus with subdivided phase structure.

Once a nucleus grows to sufficient size, it may subdivide into multiple domains. This thermodynamic process is limited by competing changes in the free energy of two sources due to the: appearance of the domain wall (which increases the energy) and the reduction of the depolarizing field due to the multi-domain structure, which decreases the overall energy of the system. Comparing the variation of the free energy originating from these sources allows for the determination of the critical radius for which the nucleus divides $\left(R_D\right)$ into multiple domains, see figure 6.

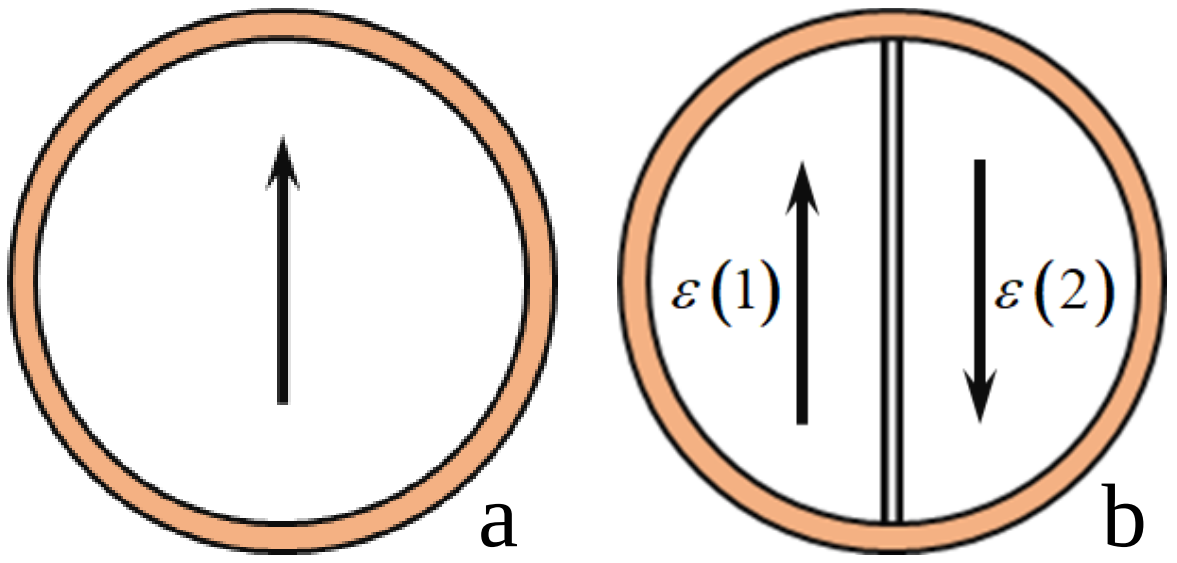

Fig.6. Comparison of nuclei structure for a single-phase (a) or subdivided nucleus (b).

A thorough analysis of the strain energy for a multi-domain nucleus having two different types of tetragonal structures, being surrounded by a cubic matrix can be found in refs. [27, 41]. The elastic energy was shown to be separated into homogeneous and heterogeneous components, with the latter describing the energy due to the internal structure made up of tetragonal domains of two kinds [27]

$$E_{Elast} = E_{Homog} + E_{Heter} ,$$

where $E_{Homog}$ is given by the expression for $\underline{U_{Elastic}}$ in (10) with $\tilde{u}_{ij}\left(\vec{k}\right) = \tilde{u}_{ij}^{0}\left(\vec{k}\right)$ corresponding to the homogeneous part of nucleus. The heterogeneous term was shown to be proportional to the internal boundary separating the differing tetragonal domains; this energy arises from short-range elastic strain concentrated inside the interphase boundary

$$E_{Heter} = \frac{1}{2}\int \frac{d^3k}{\left(2\pi\right)^3}\left(\Delta\sigma_{ij}^{0}\Delta u_{ij}^{0} - n_i\Delta\sigma_{ik}^{0}\Omega_{kl}\left(\vec{n}\right)\Delta\hat{\sigma}_{lm}^{0}n_m\right)\left|\Delta\theta\left(\vec{k}\right)\right|^2 \tag{15}$$

where $\Delta u_{ij}^0 = u_{ij}^0(1) - u_{ij}^0(2)$ and $\Delta\sigma_{ij}^0 = \sigma_{ij}^0(1) - \sigma_{ij}^0(2)$ are the difference in strains and stress in two parts of the nucleus. The interphase boundary separating the two tetragonal phase regions is commensurate with the typical thickness of the tetragonal domain wall. The shape, size and mutual arrangement of the domains are determined from the minimization of the heterogeneous energy and has been shown to be proportional to the interphase area [27]. In our model, depicted in figure 6, the tetragonal regions are of the same type with opposite direction of the polarization vector; this orientation creates the uncharged domain boundary at the center of the subdivided nucleus (figure 6.b)

$$u_{ij}^0(1) = \begin{pmatrix} u_{11}^0 & 0 & 0 \\ 0 & u_{11}^0 & 0 \\ 0 & 0 & u_{33}^0 \end{pmatrix} = u_{ij}^0(2) \ .$$

The total free energy for the inhomogeneous system (the polarization dependent part of (8)) is a function of the field variable $\vec{P}(\vec{r})$

$$F = \int \left\{ \Phi_P(P_i) + \Phi_q(P_i, u_{ij}) + g_{ijkl}\left(\frac{\partial P_i}{\partial r_j}\frac{\partial P_k}{\partial r_l}\right) - \vec{E}_{int}\cdot\vec{P}(\vec{r}) \right\} dV + \frac{1}{2}\int \frac{n_i n_j}{\varepsilon_0 \varepsilon_r} P_i(\vec{r}) P_j(\vec{r})\, d^3 r \ . \quad (16)$$

The diameter of the nucleus is considered to be much larger than the diffuse boundary around the circumference of the nucleus; therefore, the polarization appearing in the thermodynamic potential $\Phi(P_i)$ is uniform within the single-domain nucleus; this results in the appearance of a (uniform) depolarizing field opposed to the polarization vector. The energy of the single-phase nuclei is given by the following equation.

$$\Delta F_A = \left(\Phi_{Total} + E_{Dep} P_0\right) V + \int \left\{ g_{ijkl}\left(\frac{\partial P_i}{\partial r_j}\frac{\partial P_k}{\partial r_l}\right) \right\} dV + \sigma S \quad (17)$$

The polarization inside the sub-divided nucleus depends on the radius perpendicular to the domain wall (here, the *x*-axis) and is modeled by the 180° domain wall solution for the tetragonal phase [11]

$$P_z(x) = \frac{P_0 \sinh(x/\xi)}{\sqrt{A + \sinh^2(x/\xi)}} \ .$$

The energy for the sub-divided nucleus is therefore:

$$\Delta F_B = \left(\Phi_{Total}\right)V + \int \left\{ g_{ijkl}\left(\frac{\partial P_i}{\partial r_j}\frac{\partial P_k}{\partial r_l}\right) + \frac{1}{2\varepsilon_0\varepsilon_r}\left|\vec{n}\cdot\vec{P}\left(\vec{r}\right)\right|^2 \right\} dV + \sigma S + U_{180^\circ} S \ . \tag{18}$$

At a certain radius, the splitting of the nucleus becomes thermodynamically favorable and at this point the radius of these two systems is identical $\left(V_A = V_B\right)$. Moreover, the energy density of the interphase boundary shell, separating the metastable nucleus from the surrounding cubic matrix, is described by equation (7) for both configurations. The strain in the vicinity of 180° domain wall has been analyzed in [42] and the inhomogeneous strain was shown to be zero outside of the center of the domain wall. The polar component of the thermodynamic potential for the tetragonal phase is equal regardless of the presence of the domain wall, thus $\Phi_A = \Phi_B$. Both types of nuclei contain the diffuse interphase boundary separating the cubic matrix from the tetragonal nucleus; therefore, the terms given by the gradient of the polarization vector in equations (17) and (18) cancel giving the following formula.

$$\Delta F_B - \Delta F_A = \int \frac{1}{2\varepsilon_0\varepsilon_r}\left|\vec{n}\cdot\vec{P}_B\left(\vec{r}\right)\right|^2 dV + U_{180^\circ}\left(2\times 2\pi R_C^2\right) - \left(E_{Dep}P_0\right)\left(\tfrac{4}{3}\pi R_C^3\right) \ . \tag{19}$$

The inhomogeneous polarization distribution inside the sub-divided nuclei is modeled to occur as a linear combination of the diffuse interphase boundary solution (12) and the tetragonal 180° domain wall solution (2)

$$P_B\left(r\right) = \frac{1}{2}\left(\frac{\sinh\left[\left(r+R_C\right)/\xi\right]}{\sqrt{A+\cosh^2\left[\left(r+R_C\right)/\xi\right]}} + \frac{\sinh\left[\left(r+R_C\right)/\xi\right]}{\sqrt{A+\cosh^2\left[\left(r+R_C\right)/\xi\right]}}\right) - \frac{\sinh\left(r/\delta\right)}{\sqrt{A+\cosh^2\left(r/\delta\right)}} \ . \tag{20}$$

Since the integration of the first term equation (19) is over the whole volume, the deviation from the tanh($x$) profile provided by the sixth order terms in the potential is minimal. An additional assumption is that the interphase boundary width $\left(\xi\right)$ and the 180° domain wall width $\left(\delta\right)$ which divides the two-phase nucleus configuration are nearly equivalent. For this reason, the polarization distribution inside the sub-divided nucleus is given by the following formula for the integral term in (19) to be solvable

$$P_B\left(r\right) = \frac{1}{2}\left\{\tanh\left[\left(r+R_C\right)/\xi\right] + \tanh\left[\left(r-R_C\right)/\xi\right]\right\} - \tanh\left(r/\xi\right) \ . \tag{21}$$

Comparison of profiles of $P\left(r\right)/P_0$ for two solutions (20) and (21) is presented in figure 7.

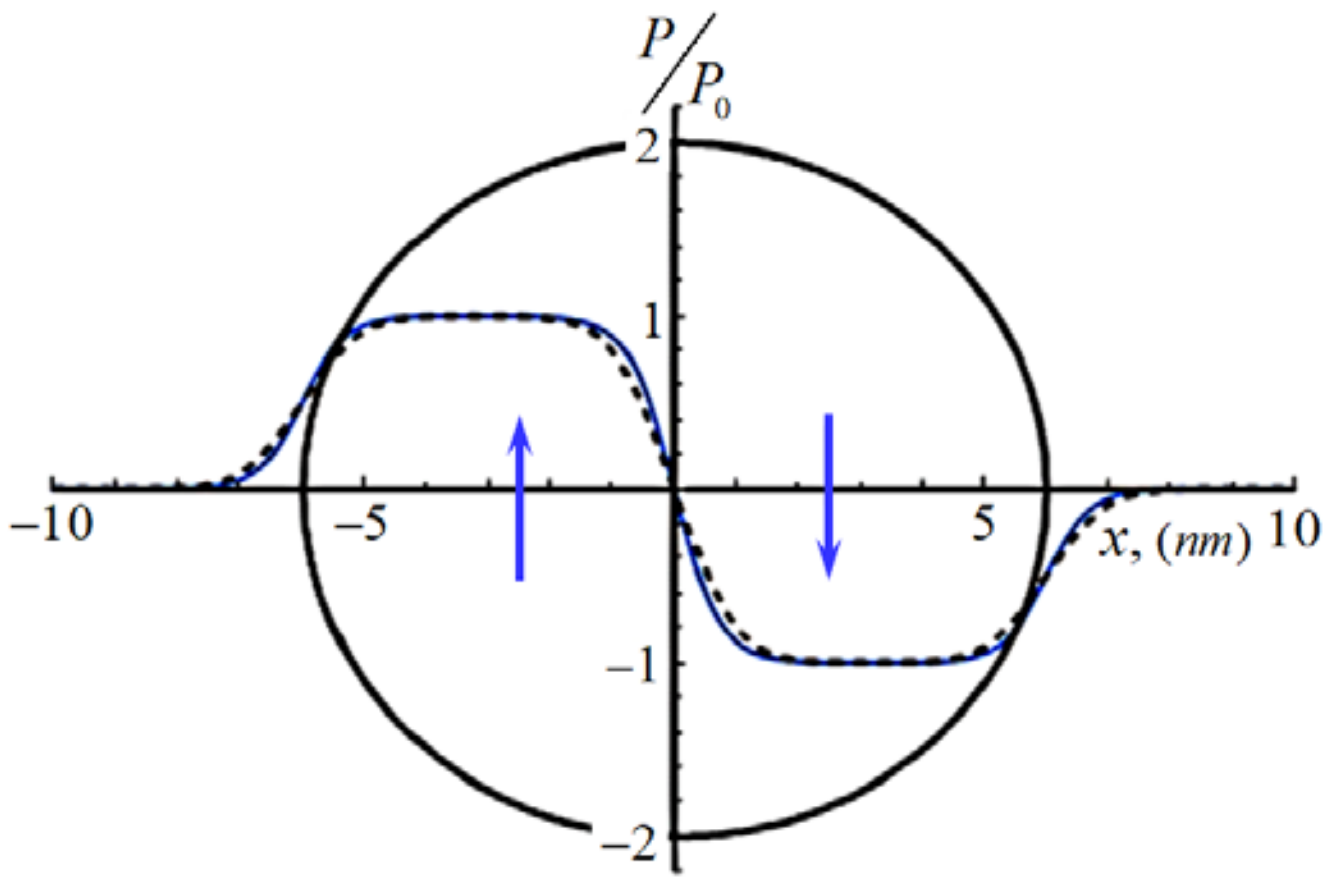

Fig. 7. Comparison of the polarization solution (sinh($x$) function) and approximation (tanh($x$)) inside the subdivided nucleus (for example $R_C$ = 6 $nm$). Arrows indicate direction of the dipole moment in domains.

The integrand in (19) is an even function, therefore, the integration limits can be changed to be symmetric about the $y$-axis

$$\frac{4\pi}{2\varepsilon_0\varepsilon_r}\int_0^{R_C}\left|\vec{P}_B\left(x\right)\right|^2 r^2 dr = \frac{1}{2}\frac{4\pi}{2\varepsilon_0\varepsilon_r}\int_{-R_C}^{R_C}\left|\vec{P}_B\left(\vec{r}\right)\right|^2 r^2 dr \ .$$

Evaluation of the above integral gives the following result

$$\frac{4\pi}{2\varepsilon_0\varepsilon_r}\int_0^{R_C}\left|\vec{P}_B\left(x\right)\right|^2 r^2 dr = \gamma\left(\frac{1}{\varepsilon_0\varepsilon_r}\right)\frac{4\pi}{3}R_C^3 \ . \tag{22}$$

The energy density stored in the single 180° domain wall is a function of temperature and has been analyzed in ref. [42]

$$U_{180^\circ} = \left(\frac{2g_{44}}{\alpha_{111}}\right)^{1/2}\left[\frac{\tilde{\alpha}_{11}}{2}P_0\sqrt{3P_0^2+\frac{\tilde{\alpha}_{11}}{\alpha_{111}}} - \left(\frac{\tilde{\alpha}_{11}^2}{4\alpha_{111}} - \alpha_1\right)\mathrm{Sinh}^{-1}\left(\frac{P_0}{\sqrt{2P_0^2+\tilde{\alpha}_{11}/\alpha_{111}}}\right)\right] , \tag{23}$$

here $\tilde{\alpha}_{11}$ is

$$\tilde{\alpha}_{11} = \alpha_{11} - \left(\frac{1}{2}q_{11}Q_{11} + q_{12}Q_{12}\right) . \tag{24}$$

As mentioned earlier, the strain inside the general ellipsoid is treated as uniform. Furthermore, coefficients used in calculations are obtained for ordered phases and thus contain the renormalization given by equation (24). Using equations (19) - (24), the difference in the free energy for the two types of nuclei are given as the following.

$$\Delta F_B - \Delta F_A \le 0 \qquad \rightarrow \qquad U_{180^\circ}\left(4\pi R_{SD}^2\right) - \left(E_{Dep}P_0 - \gamma\left(\varepsilon_0\varepsilon_r\right)^{-1}\right)\left(\tfrac{4}{3}\pi R_{SD}^3\right) \le 0 \ , \tag{25}$$

where

$$R_{SD} \geq \frac{-3\left(U_{180^\circ}\right)}{P_0 E_{Dep} - \gamma P_0^2 \left(\varepsilon_0 \varepsilon_r\right)^{-1}} \qquad \text{and} \qquad \gamma = \frac{3}{8}\left(14.6 + 8\xi\right) \approx 7.5\left(T = T_C\right) \cdot$$

For a depolarizing field having a magnitude of $10^6$ *V/m*, the critical radius for which the nucleus sub-divides into multiple domains are equal 33 *nm* (near the transition point $T_{Tr}$ = 380 *K*).

Comparing the free energy expressions for each nucleus configuration, gives the following temperature dependent expressions:

$$\Delta F_A = \left(\Phi_{Total} + E_{Dep} P_0\right)\frac{4}{3}\pi R_C^3 + \int \left\{ g_{ijkl}\left(\frac{\partial P_i}{\partial r_j}\frac{\partial P_k}{\partial r_l}\right)\right\} dV + \left(\frac{g_{44} P_0^2}{32\xi}\right) Q\left(A\right) 4\pi R_C^2 \qquad (26)$$

and after the substitution of results for σ and $U_{180^\circ}$ into (18)

$$\Delta F_B = \left(\Phi_{Total} + \frac{\gamma}{\varepsilon_0 \varepsilon_r}\right)\frac{4\pi}{3} R_C^3 + \int \left\{ g_{ijkl}\left(\frac{\partial P_i}{\partial r_j}\frac{\partial P_k}{\partial r_l}\right)\right\} dV + \left(\frac{g_{44} P_0^2}{32\xi} Q_A + U_{180^\circ}\right) 4\pi R_C^2 \quad . \qquad (27)$$

The gradient terms appearing in equations (26) and (27) are

$$\int \left\{ g_{ijkl}\left(\frac{\partial P_i}{\partial r_j}\frac{\partial P_k}{\partial r_l}\right)\right\} dV \simeq -\frac{g_{44}\xi}{4R_C^3}\mathrm{Li}_2\left(-e^{\frac{4R_C}{\xi}}\right)\frac{4\pi}{3} R_C^3 = \gamma_2 V, \qquad \gamma_2 = -\frac{g_{44}\xi}{4R_C^3}\mathrm{Li}_2\left(-e^{\frac{4R_C}{\xi}}\right) .$$

The temperature dependences of the energy density of each nuclei configuration in fig.5a and 5.b are given by the resulting formulas

$$\frac{\Delta F_B}{V} \simeq \left(\Phi_{Total} + \frac{\gamma}{\varepsilon_0 \varepsilon_r} P_0^2 + \gamma_2\right) + \frac{3}{R_C}\left(\frac{g_{44} P_0^2}{32\xi} Q_A + U_{180^\circ}\right)$$

$$\frac{\Delta F_A}{V} \simeq \left(\Phi_{Total} + \frac{3}{\varepsilon_0 \varepsilon_r} P_0^2 + \gamma_2\right) + \frac{3 g_{44} P_0^2}{32\xi R_C} Q_A$$

and are illustrated in figure 8.

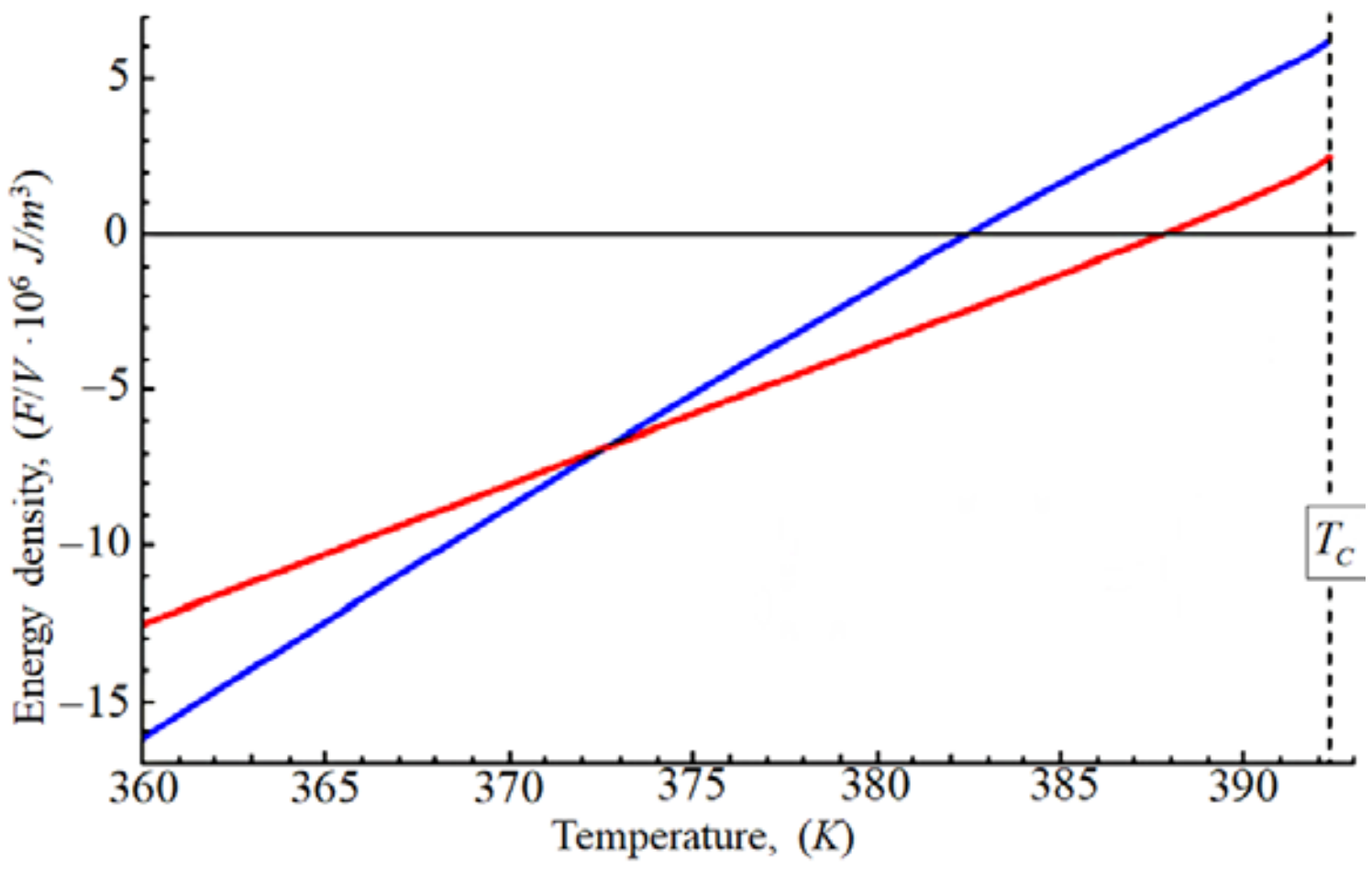

Fig. 8. Temperature dependent behavior of the free energy density for the single-phase (red line) nucleus compared to the subdivided (two-phase) nucleus (blue line).

In the process of the PE-FE phase transition, below the Curie point the tetragonal phase occurs with the single-domain nucleus having a lower free energy; thus, being the thermodynamically favorable state. As the temperature is lowered further, the magnitude of the polarization parameter continues to increase (see figure 9 given here for illustration) which leads to a larger depolarization field within the nucleus.

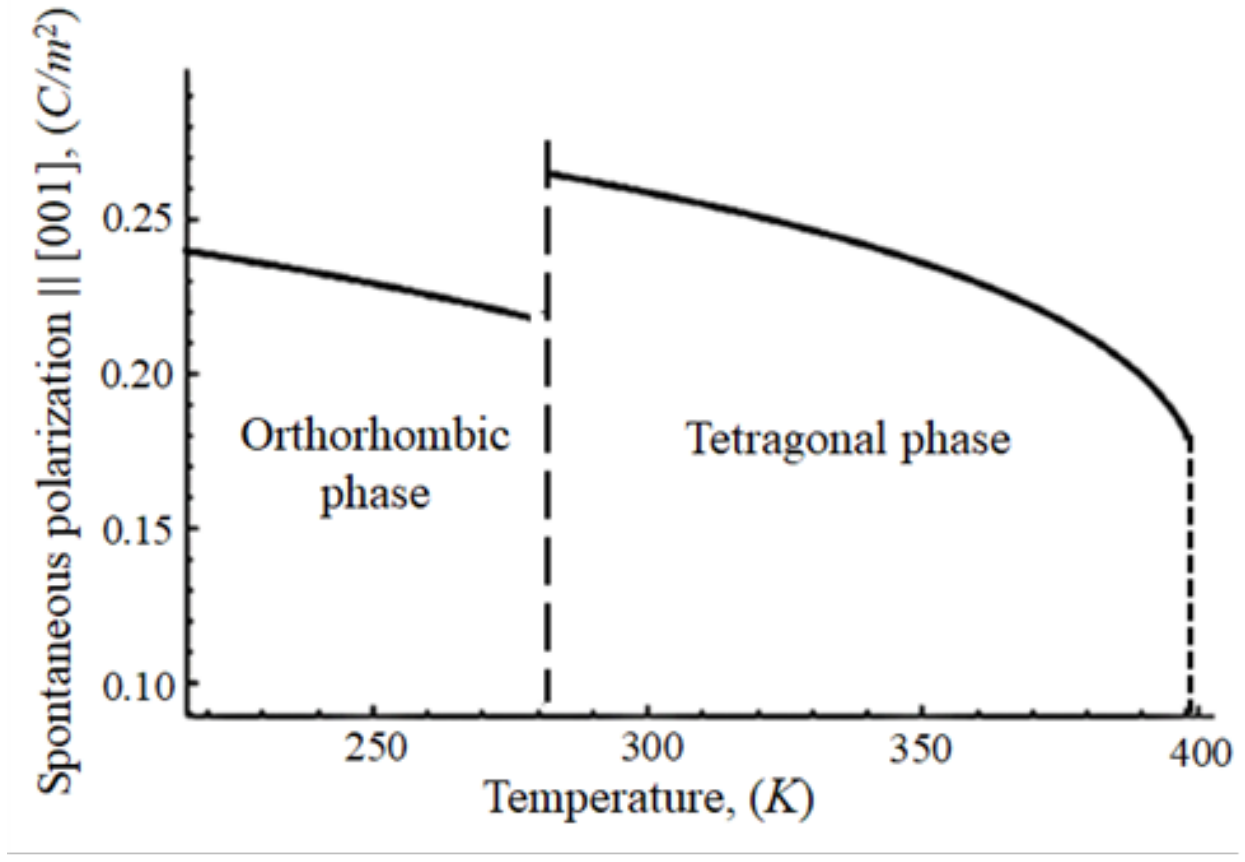

Fig. 9. Temperature dependent spontaneous polarization for the tetragonal and orthorhombic phases of $BaTiO_3$, Coefficients from ref. [22] were used for calculations with results are close to refs. [22, 43].

Meanwhile, the free energy of the subdivided nucleus is decreasing and at some temperatures becomes lower than the energy of the nucleus in single-domain state. Thus, to reduce the free energy of the growing tetragonal phase, the nucleus transitions into the subdivided configuration. It should be noted that above the tetragonal phase stability limit, the free energy functions for both nucleus types become complex and non-physical. As seen in figure 8, the free energy of both polar configurations is slightly above the free energy of the paraelectric state (*x*-axis) near the Curie point; this is due to the nucleus possessing additional terms resulting in the free energy barrier.

In conclusion we must note that the main classical assumption not addressed in our model is that the polarization near the boundary is considered planar, i.e. the nucleus curvature is not considered explicitly. Furthermore, the strain is considered uniform within the nucleus, appearing only in the thermodynamic potential. The homogeneous strain inside the ellipsoidal nucleus has already been largely discussed (see ref. [25 - 28] for example); however, the presence of long-range inhomogeneous strains near the metastable nucleus could further adjust this free energy expression. The nucleation process occurring near the surface of the sample would favor the formation of the polarized nuclei, since the surface can contain unscreened surface charges or due to the increased density of oxygen vacancies. Considering the above-mentioned features will be a future continuation of studies for the nucleation processes occurring during the ordering phase transition in ferroelectrics.

Another important point worth noting. The consideration of nuclei subdivided into domains of different phases is also of interest for studies of the ordering transition in the substances with close values of free energies of different ordered phases (for example, ferroelectric and antiferroelectric) in which the formation of two-phase nuclei has been already observed [44, 45]. The above considered type of nuclei is also possible during the phase transition from paraelectric to dipole ordered phase in compounds with compositions corresponding to morphotropic boundary region.